# Optical and Electronic Properties of SiTe$_x$ (x=1, 2) from First-Principles

*Romakanta Bhattarai and Xiao Shen*

Department of Physics and Materials Science, University of Memphis, Memphis, TN, 38152

**Abstract**

The optical and electronic properties of the α-SiTe, β-SiTe, and RX-SiTe$_2$ are investigated. A detailed analysis of electronic properties is done using standard density functional theory (DFT) and hybrid functional (HSE06) methods. The optical dielectric properties are studied under three different methods: standard DFT, many-body Green's functions (GW), and Bethe-Salpeter equation (BSE). Our calculations show that the SiTe compounds possess extremely high static dielectric constants in their bulk forms ($\varepsilon_0(\perp)$ = 68.58, $\varepsilon_0(\parallel)$ = 127.29 for α-SiTe, and $\varepsilon_0(\perp)$ = 76.23, $\varepsilon_0(\parallel)$ = 98.15 for β-SiTe). The frequency-dependent dielectric functions Im($\varepsilon$) have very large values (>100) in the optical regime, which are among the highest of layered materials, suggesting them as excellent light absorbents in the corresponding frequencies. α-SiTe exhibits a high degree of optical anisotropy as compared to the other two compounds, consistent with their structural configurations. A strong interlayer excitonic effect is observed in bulk RX-SiTe$_2$. In addition, an analysis of Raman intensity is also performed.

**Introduction**

Over the past few years, there has been a growing interest in Si-Te compounds. The study of this materials family originally started in 1953 when Weiss and Weiss synthesized SiTe$_2$ for the first time.[1] Ploog et al. synthesized Si$_2$Te$_3$ in 1976 and characterized its structure.[2] Since then, Si$_2$Te$_3$ has been the most widely studied compound in the Si-Te family,[3–11] especially after Keuleyan et al. reported in 2015 that this material can be made into 2D form.[6] A number of researches have also been done on exploring the other forms of Si-Te compounds and their properties, both theoretically and experimentally.[12–20] Like Si$_2$Te$_3$, many of them also possess 2D forms. The reduced dimension is expected to alter the electronic and optical properties of these materials and makes them potentially more interesting for optoelectronic applicantions.[21–23]

Chen et al. in 2016 proposed the α-SiTe and β-SiTe monolayers as silicon-based analogs of the black and blue phosphorene, respectively.[13] The structural, electronic, and mechanical properties of these compounds were discussed. Our recent theoretical investigation predicted a novel phase of SiTe$_2$ (which we name as RX-SiTe$_2$ here) where the Si and Te atoms exhibit a unique atomic arrangement. This phase was found to be more stable than its $CdI_2$ phase,[12] the most common phase in the IV-VI system. The optical properties of these SiTe and $SiTe_2$ compounds remain to be explored. In addition, since optical anisotropy has been observed in 2D $Si_2Te_3$,[4,5] it would be interesting for investigation in these Si-Te compounds as well.

We explore the electronic properties of the α-SiTe and β-SiTe monolayers by both the standard DFT and the hybrid density functional theory (HSE06[24,25]) methods, including the spin-orbit interactions. The analysis of the optical properties of α-SiTe, β-SiTe, and RX-$SiTe_2$ is carried out using standard DFT, many-body GW,[26–28] and BSE[29,30] methods. Our calculations show that α-SiTe and β-SiTe possess large static dielectric constants and large values of the imaginary part frequency-dependent dielectric functions Im(ε). Among the three compounds investigated, α-SiTe also shows the most optical anisotropy. In addition, a strong interlayer excitonic effect is observed in bulk RX-$SiTe_2$. Analysis of Raman spectra in SiTe shows the shifting of major Raman active modes from bulk to monolayers. These calculations can help to investigate the materials experimentally and increase the potential usefulness of the Si-Te compounds.

**Computational Methods**

The structural optimizations and the analysis of band structure and dielectric constants are performed using the VASP package.[31] The projected augmented wave (PAW) pseudopotential[32] with exchange and correlation functional under generalized gradient approximation in the Perdew-Burke-Ernzerhof (PBE)[33] form is used in the calculations. A plane-wave basis set with the kinetic energy cutoff of 500 eV is used for the expansion of the electronic wave functions. The electronic and force convergence criteria during the structural optimization are set to $10^{-9}$ eV and $10^{-4}$ eV/Å, respectively. The Brillouin zones are sampled in the Gamma centered k-point grids of 13×13×13 for α-SiTe and β-SiTe as well as 13×13×7 for RX-$SiTe_2$ in the bulks, and 13×13×1 in the monolayers. The Raman spectra calculations are done using the Quantum ESPRESSO[34] package under the density functional perturbation theory (DFPT).[35] The norm-conserving[36] pseudo-potentials generated via Rappe-Rabe-Kaxiras-Joannopoulos (RRKJ)[37] method are used. A plane wave basis with the cutoff energy of 80 Rydberg (Ry) is used in the calculation. The total energy convergence criterion for the self-consistent calculation is set $10^{-12}$ Ry after the structures are fully relaxed under the ionic minimization conditions, where the energy and force convergences are set to $10^{-8}$ Ry and $10^{-7}$ Ry/Bohr respectively. During the phonon calculation, the convergence criterion is set to $10^{-14}$ Ry.

**Results and Discussions**

**A.     Crystal structures**

The crystal structures of α-SiTe and β-SiTe monolayers are taken from the literature.[13] After the monolayers are optimized, a reverse technique is followed to obtain the corresponding bulk structures. The resulting bulk lattice parameters are summarized in Table I. The corresponding crystal structures of α-and β-SiTe are shown in Figure 1, along with the RX-SiTe$_2$ structure. The in-plane lattice parameters of the bulk structures differ from their respective monolayers. For the α-SiTe monolayer, the optimized lattice parameters are a = 4.29 Å and b = 4.11 Å. In the bulk form, we find a = 4.077 Å, and b = 4.231 Å, which are 4.96 % smaller and 2.94 % larger than their monolayers, respectively. Also, the α-SiTe is found to preserve its orthorhombic structural symmetry from bulk to monolayer. For the β-SiTe, lattice parameters are a = b = 3.83 Å in the monolayer, which are increased by 7.07 % to a = b = 4.101 Å in the bulk form. The energies per atom of both SiTe compounds in bulk are -4.081 eV and -4.083 eV for α-SiTe and β-SiTe, respectively. Meanwhile, the energies are -4.025 eV and -3.996 eV in the corresponding monolayers, which shows ~29 meV per atom energy difference. This result is consistent with the previous finding that the monolayer α-SiTe is energetically more favorable than the monolayer β-SiTe.[13]

Table I. Lattice parameters of α-SiTe and β-SiTe in their bulk

| Phases | a (Å) | b (Å) | c (Å) | α (°) | β (°) | γ (°) | Bravais lattice |
|---|---|---|---|---|---|---|---|
| α -SiTe | 4.077 | 4.231 | 5.849 | 90 | 90 | 90 | Orthorhombic |
| β -SiTe | 4.101 | 4.101 | 4.223 | 60.949 | 119.051 | 119.999 | Triclinic |

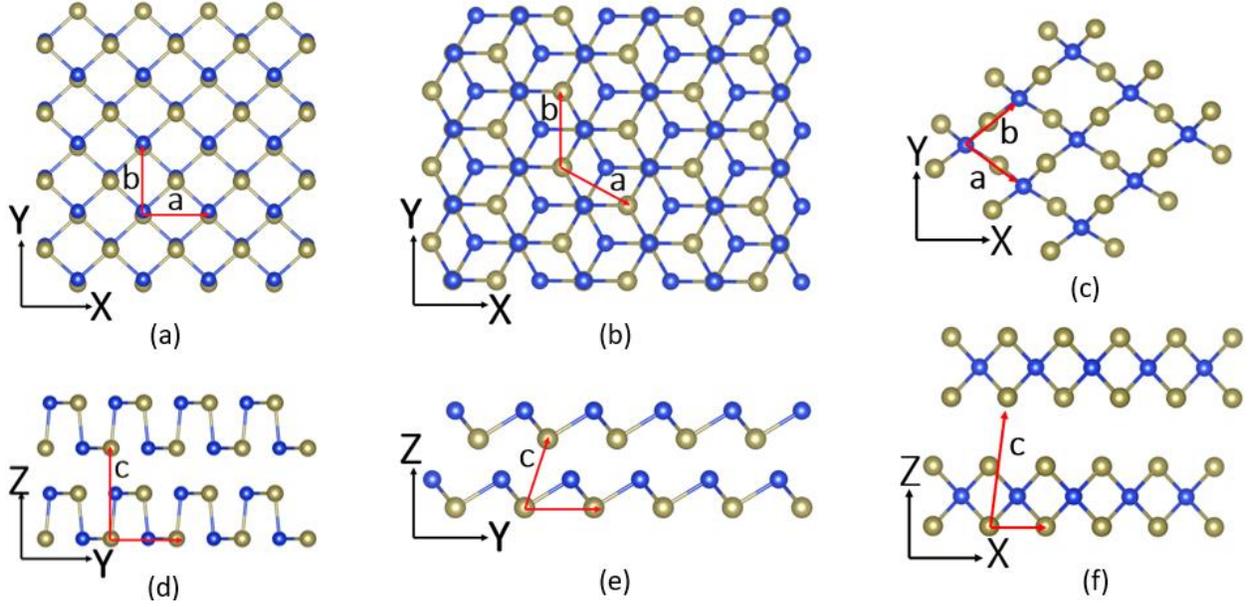

Figure 1. Crystal structures of α-SiTe, β-SiTe, and RX-SiTe$_2$ with the top views and side views represented by (a-c) and (d-f) respectively. The solid red lines indicate the lattice vectors. Si and Te atoms are represented by blue and light tan color, respectively.

**B. Electronic Properties**

To understand the electronic properties of SiTe phases, we calculate their band structures in the bulks as well as in the monolayers. The calculations are done under DFT and HSE06 method with and without taking the spin-orbit coupling (SOC) effect into account. The band structures of bulk SiTe phases with SOC are presented here (see Supplementary Material for band structures of monolayers). Figure 2 shows the electronic band structures of bulk α-SiTe with SOC, in which the DFT band structures (Figure 2 (a)) feature a slight overlapping of the valence band and conduction band, giving rise to no band gap. Figure 2 (b) is the band structures under the HSE06 method, which in most cases overcomes the limitation of standard DFT results in underestimating the band gap, thereby providing better agreement with the experiment. Under this method, the band gap is found to be 0.09 eV and is indirect in nature. This value is 0.15 eV if we neglect the SOC effect in the calculations.

Table II. Band gaps of SiTe compounds under DFT and HSE06 methods

|  |  | DFT |  | HSE06 |  |
| --- | --- | --- | --- | --- | --- |
| SiTe Phases |  | Without SOC | With SOC | Without SOC | With SOC |
| α -SiTe | Bulk | 0 | 0 | 0.15 | 0.09 |

|        | Monolayer | 0.40 | 0.36 | 0.67 | 0.61 |
|--------|-----------|------|------|------|------|
| β-SiTe | Bulk      | 0    | 0    | 0.23 | 0.18 |
|        | Monolayer | 1.83 | 1.54 | 2.49 | 2.20 |

Similarly, the band structures of bulk β-SiTe under DFT and HSE06 methods with SOC are shown in Figure 3. The HSE06 band gap is 0.18 eV, which becomes 0.23 eV without SOC. For the monolayers, the band gaps under DFT are 0.40 eV and 1.83 eV for the α-SiTe and β-SiTe, whereas 0.67 eV and 2.49 eV under HSE06 method without SOC. These results agree with the previously reported values.[13] However, when the SOC is turned on, the respective band gaps become 0.36 eV & 1.54 eV under DFT as well as 0.61 eV & 2.20 eV under HSE06 method. This implies that the spin-orbit interaction is lowering the band gap by 0.04-0.06 eV in α-SiTe monolayer and 0.29 eV in β-SiTe monolayer, thus suggesting the presence of strong SOC effect in β-SiTe. The band gaps of the two SiTe phases from different methods are presented in Table II.

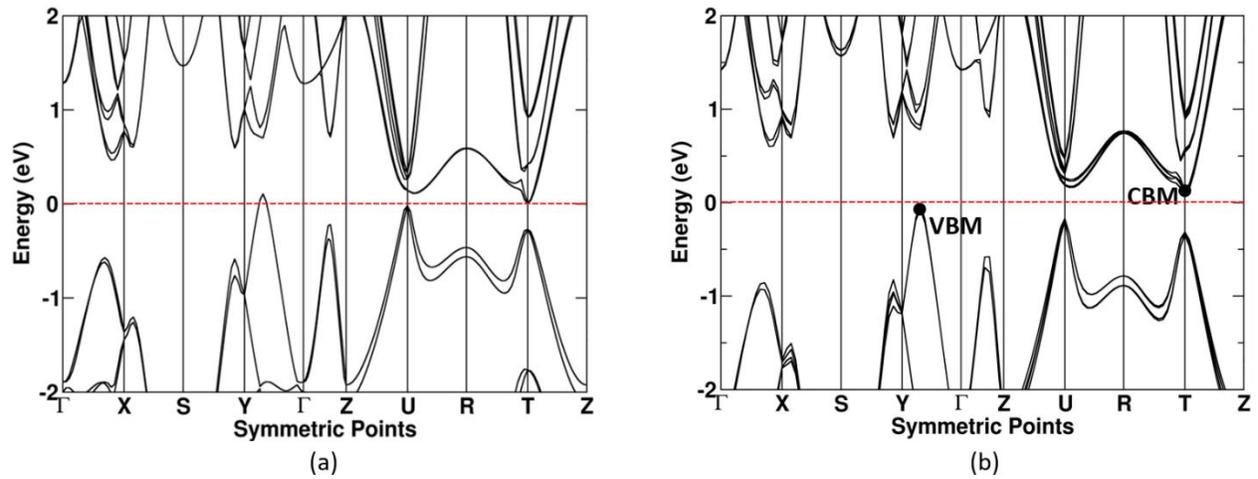

Figure 2. Electronic band structures of bulk α-SiTe from (a) DFT (b) HSE06 method. The Fermi level is represented by dotted red line.

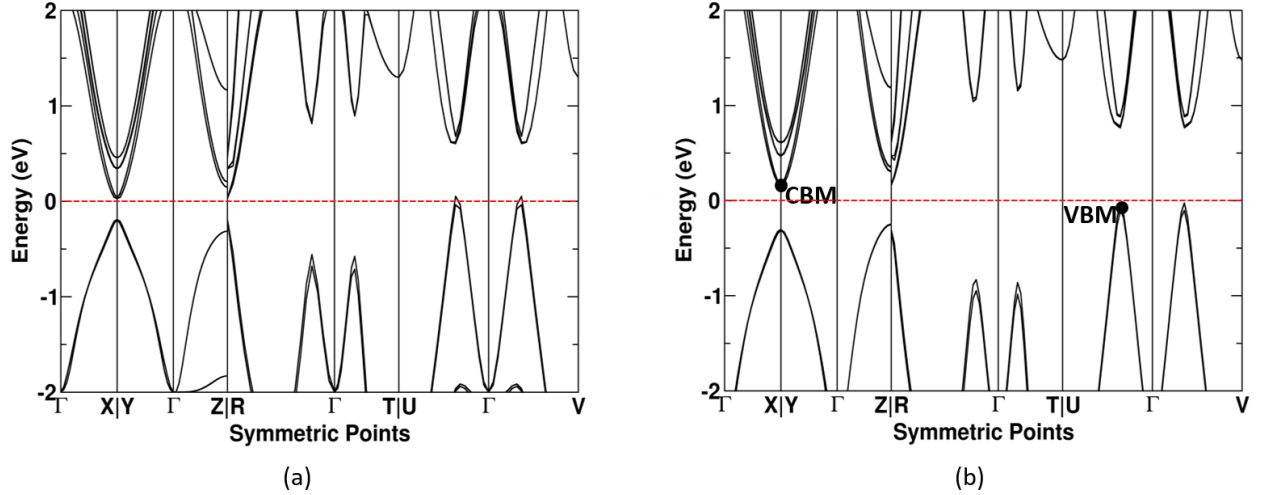

Figure 3. Electronic band structures of bulk β-SiTe from (a) DFT (b) HSE06 method.

### C. Dielectric Properties

Next, we discuss the static dielectric constants of the three compounds under the DFT method. In the bulk, the calculated dielectric constants for α-SiTe are $\varepsilon_0(x) = 75.23$, $\varepsilon_0(y) = 61.92$, and $\varepsilon_0(z) = 127.29$, while for β-SiTe, these values are 76.23, 76.23, and 98.15, respectively. Taking an average, we find that $\varepsilon_0(\perp) = 68.5$, $\varepsilon_0(\parallel) = 127.29$ for α-SiTe and $\varepsilon_0(\perp) = 76.23$, $\varepsilon_0(\parallel) = 98.15$ for β-SiTe. These values are much larger than the dielectric constants of the $Si_2Te_3$.[4] Although the large dielectric constants in the bulk, their monolayers exhibit relatively small values, being $\varepsilon_0(x) = 23.36$, and $\varepsilon_0(y) = 18.31$ for α-SiTe as well as $\varepsilon_0(x) = \varepsilon_0(y) = 3.96$ for β-SiTe. Clearly, the in-plane dielectric function is anisotropic in α-SiTe and is isotropic in β-SiTe. The structural configuration appears to be the major cause of this difference. Since the geometrical configurations of α-SiTe and β-SiTe are similar with black phosphorene and blue phosphorene, respectively. α-SiTe exhibits armchair and zigzag structures, which in this case are x-axis and y-axis, respectively. As a result, α-SiTe has anisotropic static dielectric constants along x- and y-directions. However, β-SiTe has a higher degree of structural symmetry and no anisotropy in the x-y plane, therefore exhibit the same values of $\varepsilon_0(x)$ and $\varepsilon_0(y)$.

The static dielectric constants of the bulk RX-$SiTe_2$ are $\varepsilon_0(x) = 12.91$, $\varepsilon_0(y) = 11.28$, and $\varepsilon_0(z) = 6.10$, whereas for the monolayer, they are $\varepsilon_0(x) = 5.55$, and $\varepsilon_0(y) = 4.93$. Taking an average, we find that $\varepsilon_0(\perp) = 12.09$, $\varepsilon_0(\parallel) = 6.10$ for bulk $SiTe_2$. The low degree of in-plane structural asymmetry in RX-$SiTe_2$ is the reason behind the similar dielectric constants in the in-plane directions. These values of $\varepsilon_0$ in RX-$SiTe_2$ are significantly lower than those of the SiTe compounds.

### D. Optical Properties

Here we discuss the frequency-dependent dielectric functions. Figures 4, 5, and 6 show Im(ε), the imaginary parts of the dielectric functions of α-SiTe, β-SiTe, and RX-SiTe$_2$ as a function of corresponding photon energies. From Figure 4, one can see that Im(ε) in bulk α-SiTe is anisotropic along the x- and y-directions, as in the case of the static dielectric constants. It is also observed that Im(ε) along the out-of-plane direction (z-axis) is much larger than the in-plane direction in bulk α-SiTe, which is the opposite in the case of bulk β-SiTe (Figure 5), where Im(ε) along z-axis is lower. The in-plane optical spectra in β-SiTe are found similar in x- and y-directions, which is also similar to the behavior of the static dielectric constants discussed above. From the RX-SiTe$_2$ (Figure 6), the dielectric function in the x- and y-directions are close, again consistent with the behavior of the static dielectric values, which is due to the less asymmetric nature of the crystal structure. One interesting feature of the Im(ε) in RX-SiTe$_2$ is that the BSE method yields a very different result for the Im(ε) along the z-axis compared with the DFT and GW results, thus indicating a strong inter-plane excitonic effect in RX-SiTe$_2$, similar to the case of bulk Si$_2$Te$_3$.[4]

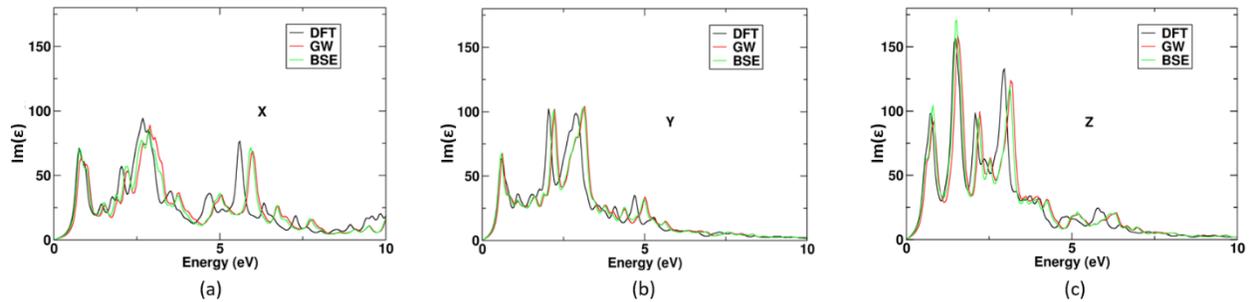

Figure 4. Imaginary part of frequency dependent dielectric function of bulk α-SiTe.

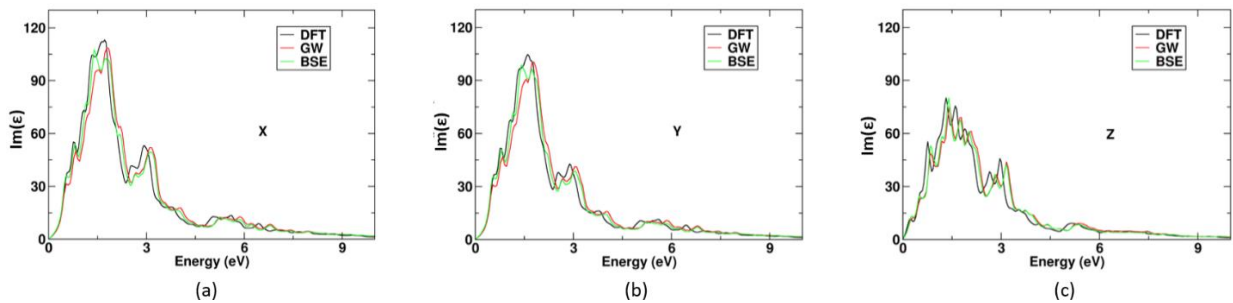

Figure 5. Imaginary part of frequency dependent dielectric function of bulk β-SiTe.

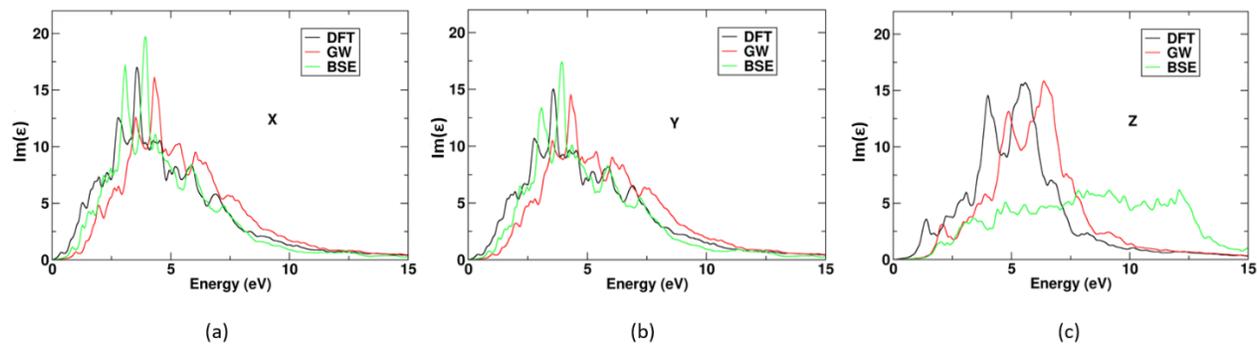

Figure 6. Imaginary part of frequency dependent dielectric function of bulk RX-SiTe$_2$

From the GW calculations, we obtain the quasiparticles (QP) band gaps of 0.05 eV, 0.12 eV, and 1.16 eV for the bulk α-SiTe, β-SiTe, and RX-SiTe$_2$, respectively. The corresponding QP band gaps for the monolayers are calculated as 0.50 eV, 2.93 eV, and 2.03 eV. As expected, these GW band gaps are larger than their DFT counterparts. By comparing the dielectric function curves from GW and BSE calculations, we find the exciton binding energies in the bulk α-SiTe, β-SiTe, and RX-SiTe$_2$ as 0.026 eV, 0.019 eV, and 0.133 eV, respectively. This indicates that the excitons are loosely bound in these bulk α- and β-SiTe but are strongly bound in bulk RX-SiTe$_2$. The latter result is consistent with the behavior of Im(ε) along the z-axis in RX-SiTe$_2$ (Figure 6c), where the BSE calculation introduces a significant modification of the spectra, indicating a strong interlayer exciton effect. We also calculate the exciton binding energies in the respective monolayers (see Supplementary Material). From Figure S2 in the Supplementary Material, the exciton binding energies for the monolayers are calculated as 0.122 eV, 0.651 eV, and 0.455 eV. These results suggest the large excitonic binding energies in the monolayer β-SiTe and RX-SiTe$_2$, but not in the α-SiTe monolayer.

To gain insights into the large exciton binding energy in bulk RX-SiTe$_2$ and the large change of Im(ε) in the out-of-plane direction under the BSE method, we plot the hole and electron charge densities of RX-SiTe$_2$ at the HOMO (highest occupied molecular orbital) and LUMO (lowest unoccupied molecular orbital) in Figure 7. Only the side views are shown to facilitate the analysis of charge densities along the z-axis. From Figure 7, no orbital contribution is found from the Si atoms (marked 1) in the HOMO. Instead, only the 5p-orbitals of Te atoms (marked 2) are contributing. Meanwhile, the LUMO consists of 2s-orbitals of Si (marked 3) as well as 5p-orbitals of Te atoms (marked 4). These results show that the p-orbitals of Te atoms contributes to both HOMO and LUMO and extend into the space between the layers in RX-SiTe$_2$ similar to Si$_2$Te$_3$.[4] This enables close proximity between the electrons and holes in the two adjacent layers, hence causing the strong interlayer excitonic effect.

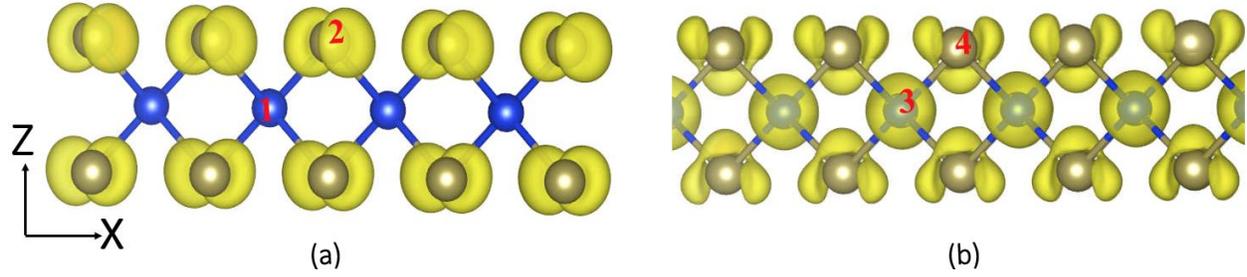

Figure 7. Charge density plot of RX-SiTe$_2$ side view at (a) HOMO and (b) LUMO. Iso-surface levels are 0.0105 and 0.0086, respectively.

It is important to note that α- and β-SiTe phases have very large values of Im(ε) in the optical regime. From Figure 4 in α-SiTe, the Im(ε) from the BSE approach has a maximum intensity of 85 at 2.88 eV along the x-axis, which is 103 at 3.12 eV along the y-axis, and 171 at 1.52 eV along the z-axis. Similar behavior can be observed under the DFT and GW approaches. Similarly, from Figure 5 in β-SiTe, the Im(ε) from the BSE approach has a maximum of 107, 98, and 80 at the photon energy of 1.40 eV along the x-, y-, and z-axes, respectively. These values are among the highest of layered materials.[4,38–53] Overall, the findings suggest that α- and β-SiTe phases can be excellent light absorbents in the corresponding frequencies.

To help further study SiTe phases, we calculate their Raman spectra under density functional perturbation theory. Figure 8 represents the Raman spectra of α-SiTe, and β-SiTe. The Raman spectra of RX-SiTe$_2$ are discussed in Reference 12. Figure 8 (a) shows that bulk α-SiTe have major Raman active peaks at 137 cm$^{-1}$, and 194 cm$^{-1}$, along with a very weak peak at 154 cm$^{-1}$ (not seen in Figure). For the monolayer, there are three peaks at 133 cm$^{-1}$, 181 cm$^{-1}$, and 252 cm$^{-1}$. The Raman peaks for β-SiTe from Figure 8 (b) appear at 124 cm$^{-1}$ and 198 cm$^{-1}$ in the bulk and at 245 cm$^{-1}$ and 334 cm$^{-1}$ in the monolayer. These large shifts indicate a strong effect of interlayer coupling in β-SiTe. These results may be useful for the experimental study of these materials.

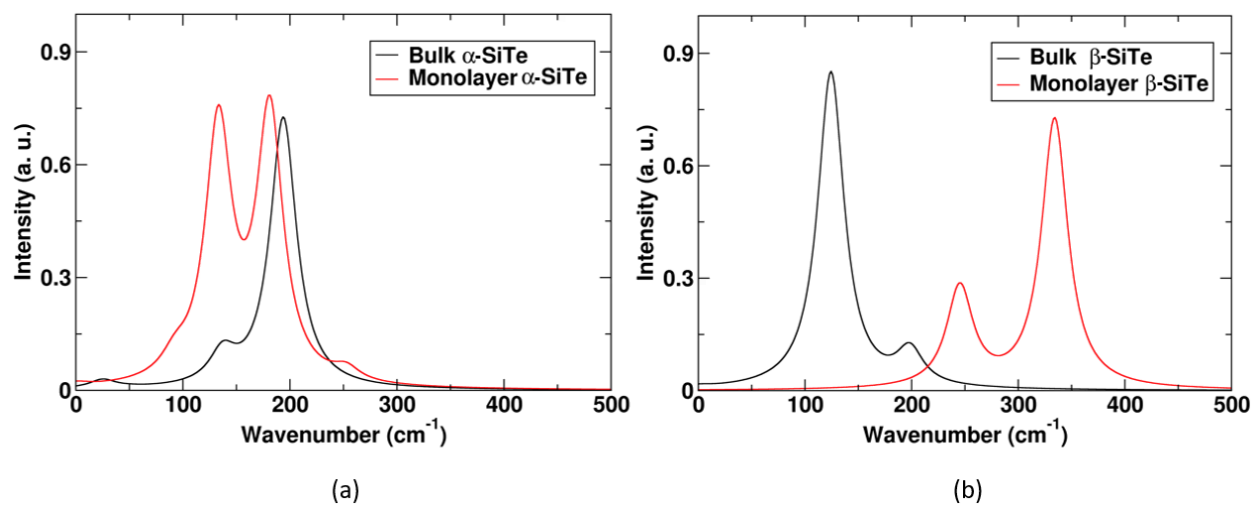

Figure 8. Raman spectra (a) α-SiTe (b) β-SiTe

**Conclusions**

The electronic and optical properties of SiTe$_x$ (x=1, 2) compounds are explored. The comparative band analysis of the α-SiTe, β-SiTe, and RX-SiTe$_2$ under different approaches is presented. The optical properties of the compounds are calculated under three major approximations: DFT, GW, and BSE methods. Large static dielectric constants are found in α- and β-SiTe. The frequency-dependent dielectric function shows the high degree of optical anisotropy in α-SiTe as compared to β-SiTe and RX-SiTe$_2$, which is due to the anisotropic geometrical configuration of α-SiTe. The large excitonic effect is observed in the bulk RX-SiTe$_2$ along the out-of-the plane direction. Both α- and β-SiTe possess very large Im(ε) in the optical regime, indicating they are good light absorbent. The Raman spectra are also predicted to aid further experimental investigations. Raman spectra in β-SiTe show significant changes in the frequencies and intensities of major Raman active modes from bulk to monolayers. These calculations in overall increase the potential for applications of the Si-Te compounds.

**Supplementary Material**

See Supplementary Material for the band structures and frequency dependent dielectric functions of SiTe$_x$ (x=1, 2) monolayers.

**Acknowledgement**

This work was supported by the National Science Foundation under Grant No. DMR 1709528 and by the Ralph E. Powe Jr. Faculty Enhancement Awards from Oak Ridge Associated Universities (ORAU). Computational resources were provided by the University of Memphis High-Performance Computing Center (HPCC) and by the NSF XSEDE under Grant Nos. TG-DMR 170064 and 170076. We thank Dr. T. B. Hoang for the helpful discussion.

**Data Availability**

The data that support the findings of this study are available from the corresponding author upon reasonable request.

# Supplementary Material

**Optoelectronic properties of SiTe$_x$ (x = 1, 2) from First-Principles Calculations**

*Romakanta Bhattarai and Xiao Shen*

Department of Physics and Materials Science, University of Memphis, Memphis, TN, 38152

## I.   Band Structures of α- and β-SiTe Monolayers

The band structures of α-SiTe and β-SiTe monolayers under the DFT method with spin-orbit interactions are shown in Figure S1. VBM and CBM refer to the valence band maximum and the conduction band minimum, respectively. In both the panels in Figure S1, VBM and CBM are located at two different sites in the Brillouin zone, indicating the indirect band gaps similar to their bulk cases. The steep slopes in the bands along the Γ-Y direction near VBM as well as along the Γ-X direction near CBM in α-SiTe signify the small effective masses of holes and electrons, hence high carrier mobilities. Meanwhile, relative flat bands are observed near the VBM and CBM in β-SiTe, indicating large effective masses and lower mobilities.

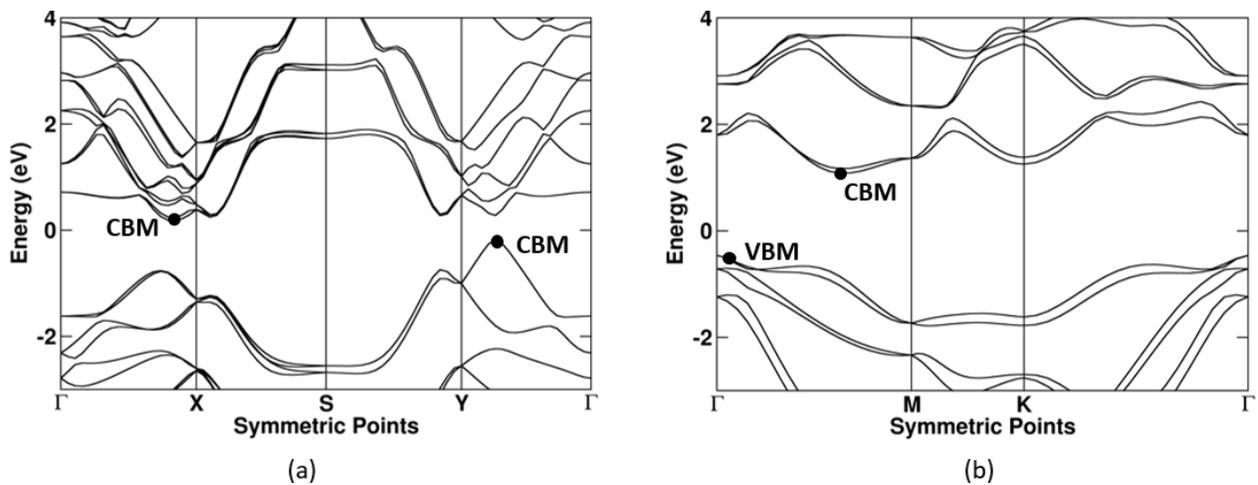

Figure S1. Electronic band structures of monolayers under DFT method (a) α-SiTe (b) β-SiTe.

## II. Frequency-Dependent Dielectric Functions of Monolayers

Figure S2, S3, and S4 show the imaginary parts of the frequency-dependent dielectric functions Im($\varepsilon$) of α-SiTe, β-SiTe, and RX-SiTe$_2$, respectively. In Figure S2, Im($\varepsilon$) along the x-axis is greater than the y-axis. The shapes of the curves are also different. The maximum intensity of the dielectric curve from the BSE method is 38 at 1.12 eV along the x-axis, and 32 at 1.80 eV along the y-axis. Similar nature is observed in DFT and GW results, meaning that α-SiTe has anisotropic optical properties in the monolayer as well. Meanwhile, no anisotropy is observed in the β-SiTe monolayer (Figure S3), where the corresponding peak value of Im($\varepsilon$) is 17 at 3.28 eV along both axes. This result is in agreement with the DFT and GW curves. In Figure S4, the dielectric function of RX-SiTe$_2$ is found to be slightly anisotropic, where the peak values of Im($\varepsilon$) along the x- and y-axis are 10 and 8 at 3.96 eV and 4.00 eV, respectively. The BSE curve exhibits an intense peak in the β-SiTe and RX-SiTe$_2$ monolayers, thus showing the significant excitonic response in these materials.

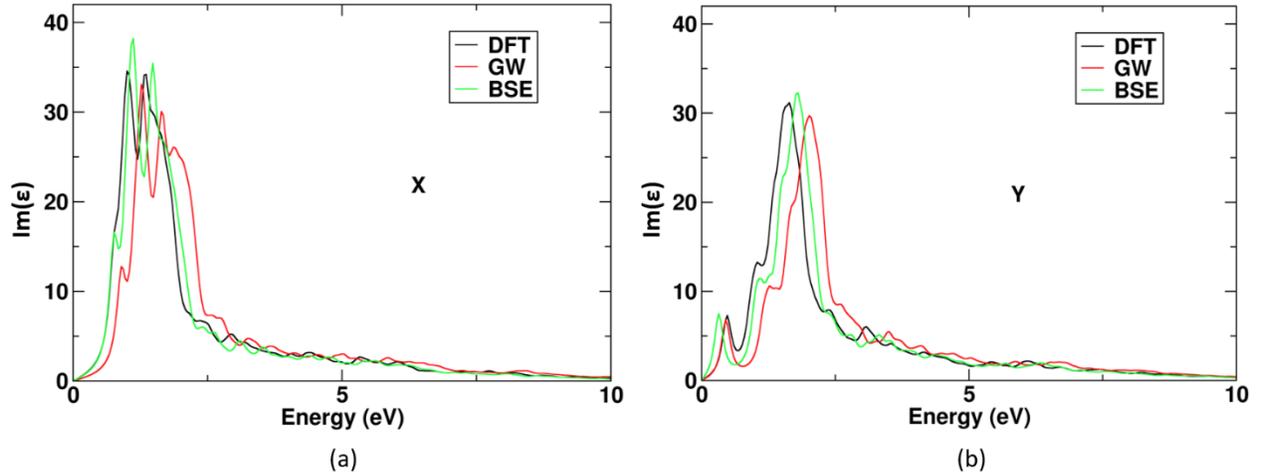

Figure S2. Imaginary part of the frequency dependent dielectric function of monolayer α-SiTe.

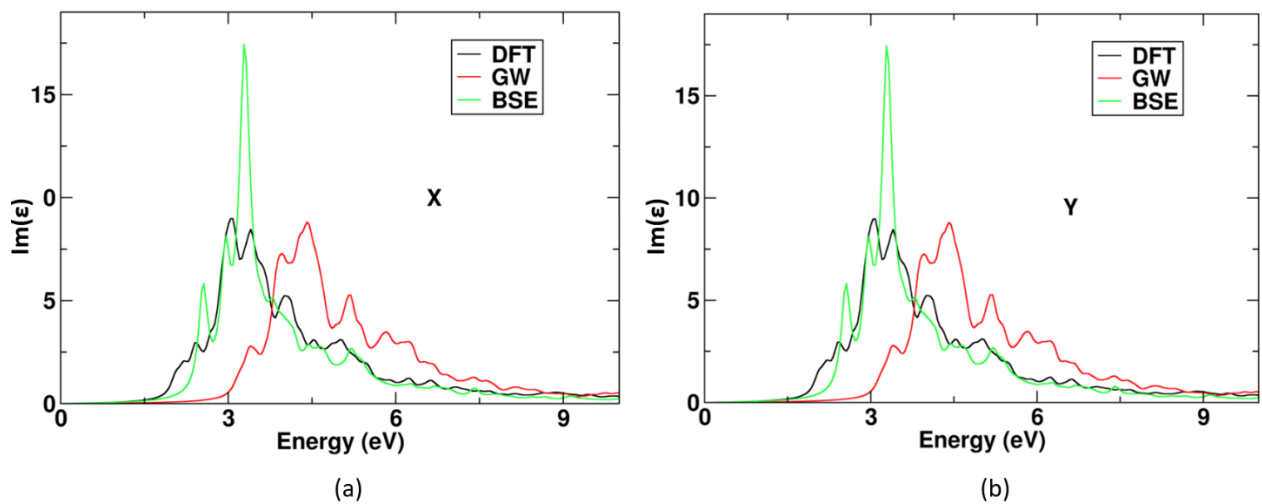

Figure S3. Imaginary part of the frequency-dependent dielectric function of monolayer β-SiTe.

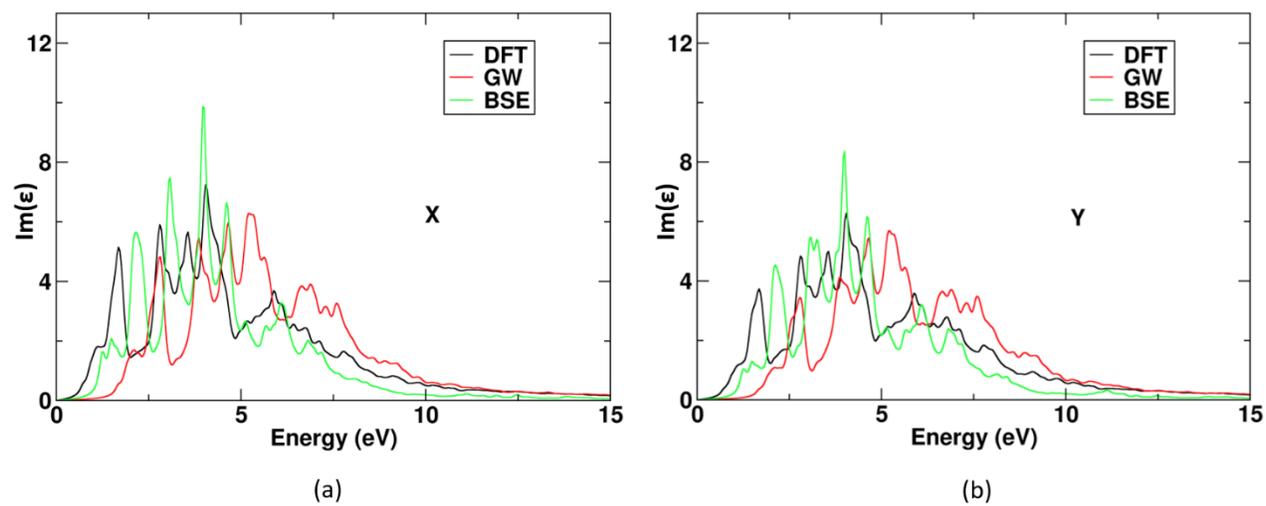

Figure S4. Imaginary part of the frequency-dependent dielectric function of monolayer RX-SiTe$_2$.